\documentclass[]{gmaart2022}

\renewcommand{\varProcName}{Proc. 34. Workshop Computational Intelligence}
\renewcommand{\varLocation}{Berlin}
\renewcommand{\varDate}{21.-22.11.2024}

\usepackage[utf8]{inputenc} 
\usepackage[T1]{fontenc} 
 
\usepackage{amsmath,amssymb,amsfonts,mathtools,amsthm} 

\usepackage[babel,autostyle]{csquotes}

\usepackage{tabularx}
\newcolumntype{L}[1]{>{\raggedright\arraybackslash}p{#1}} 
\newcolumntype{C}[1]{>{\centering\arraybackslash}p{#1}} 
\newcolumntype{R}[1]{>{\raggedleft\arraybackslash}p{#1}} 
\usepackage{booktabs}

\usepackage{paralist}   

\usepackage[noend]{algpseudocode}

\algrenewcommand\algorithmicrequire{\textbf{Voraussetzung:}}
\algrenewcommand\algorithmicensure{\textbf{Abschlussbedingung:}}

\usepackage{listings} 
\usepackage{subcaption}  

\usepackage{epstopdf}
\usepackage{xcolor}
\selectcolormodel{gray}  

\usepackage{scrhack}
\usepackage{varwidth}
\usepackage{rotating} 
\usepackage[defaultlines=2,all]{nowidow}  
\usepackage{import}
\usepackage{placeins}
\usepackage{makecell}

\usepackage{graphicx}
\usepackage{subcaption}
\usepackage{tikz}
\usepackage{pgfplots}


\usepackage{amsmath,amssymb,amsfonts,mathtools,amsthm} 

\usepackage{longtable} 
\usepackage{multirow}
\usepackage{tikz}
\usepackage{pgfplots}
\usepackage{smartdiagram}
\usetikzlibrary{patterns}
\usetikzlibrary{shapes} 
\usetikzlibrary{shapes.geometric, arrows,positioning}
\usepackage{smartdiagram}
\usesmartdiagramlibrary{additions} 
\AtBeginDocument{
}
\usepackage{xcolor}
\selectcolormodel{gray}

\usepackage[acronym]{glossaries} 
\usepackage{hyphenat}
\usepackage{etoolbox}
\usepackage{printlen}
\usepackage{graphicx}
\usetikzlibrary{calc, positioning, matrix}
\usepackage{standalone}

\newcommand{\newacr}[4][]{\newacronym[
	sort={\ifthenelse{\isempty{#1}}{#2}{#1}},
	]{#2}{#3}{#4}}
\glsdisablehyper

\robustify{\bfseries}

\begin{document}


\hyphenpenalty=2000

\pagenumbering{roman}
\tableofcontents  
\cleardoublepage
\setcounter{page}{1}
\pagestyle{scrheadings}
\pagenumbering{arabic}

\setnowidow[2]
\setnoclub[2]

\renewcommand{\Title}{Optimized Excitation Signal Design Employing Receding Horizon Control}

\renewcommand{\Authors}{Max Heinz Herkersdorf\textsuperscript{1}, Oliver Nelles\textsuperscript{2}}
\renewcommand{\Affiliations}{University of Siegen\\
	Paul-Bonatz-Straße 9-11, 57076 Siegen\\
	\textsuperscript{1}E-Mail: max.herkersdorf@uni-siegen.de\\
	\textsuperscript{2}
		E-Mail: oliver.nelles@uni-siegen.de}

							 
\renewcommand{\AuthorsTOC}{M. Herkersdorf, O. Nelles} 
\renewcommand{\AffiliationsTOC}{University of Siegen} 

\setLanguageEnglish
							 
\setupPaper 


\section*{Introduction}
Contemporary nonlinear system identification applications leverage powerful machine learning techniques to a great extent. The effectiveness of these data-driven approaches is significantly influenced by the quality of the input or excitation signals employed to generate training and validation datasets. Consequently, alongside the selection of an appropriate model architecture and parameter estimation strategy, the methodologies for input signal design are of paramount importance. \\
\\
The fundamental objective of input signal design methodologies is to obtain precise and comprehensive information concerning the process behaviors intended to be modeled. Especially considering real-world limitations like time and process constraints as well as high measurement costs \cite{rivera2009constrained}, designing input signals becomes a highly application-specific challenge. This challenge inherently involves a tradeoff between acquiring information in unknown operational areas (exploration) and refining knowledge in established areas (exploitation).

This contribution presents a novel strategy for generating excitation signals tailored to nonlinear dynamic processes. Drawing inspiration from receding horizon control (RHC), the excitation signal is optimized in an iterative manner, with each iteration focusing solely on a finite time horizon. In the optimization process, a newly introduced criterion is pursued, enabling user-defined adaptations of information acquisition to align with relevant process behaviors while also accommodating real-world limitations. This flexibility enables the method to respond effectively to application-specific challenges.

The proposed approach aligns with a recent research field focused on optimizing the distribution within the input space of a nonlinear dynamic process \cite{herkersdorf2024idsfid,kiss2024space,heinz2017iterative}. These approaches are grounded in the principle that the information collected by an excitation signal about a process is inherently linked to its generated distribution within the process's input space. Consequently, the challenge of "gathering information about the process behavior intended to be modeled" can be reframed as the task of "exciting the relevant regions in the process's input space". The underlying assumption is that, if all relevant regions of the input space are sufficiently represented in the training (and validation) datasets, a model can be trained that will be able to accurately describe the process behavior of interest. This is particularly applicable to Markovian processes, which can be fully predicted given knowledge of the current process states \cite{van1992stochastic}.

\section*{Excitation Signal Design Strategy}
In this section, the proposed excitation signal design methodology is presented, emphasizing its key contributions: (i) the development of a RHC-like iterative approach, drawing inspiration from \cite{gedlu2023online} and (ii) the introduction of a novel optimization criterion, calculated in the input space of a nonlinear dynamic model and designed to flexibly respond to application-specific challenges.

\subsubsection*{Receding Horizon Control-Like Algorithm}
The main idea of the RHC-like excitation signal design is rooted in its iterative optimization, which is performed only within the finite time horizon $L$ each iteration.  As optimized data points $\underline{\hat{U}}_{\mathrm{opt}}$ become available through the optimization (cf.$\ $Eq.$\ $\ref{eq:RHClikeOptimization}), the optimal data point $\underline{\hat{u}}_{\mathrm{opt}}(k)$ is appended to the existing signal $\underline{{U}}$. The time horizon is then shifted forward and the optimization process is repeated until the entire signal is designed. Constraining the controllable inputs and the model input space through $\mathcal{U}$ and $\mathcal{X}$, respectively, is highly advantageous for real-world applications \cite{kosters2022optimization}.

The core aim of the optimization is to generate data points in the relevant regions of the process's input space, thereby facilitating the collection of information about its behavior intended to be modeled.  Typically, prior knowledge regarding the process is however limited, rendering direct access to the input space infeasible. Consequently, a surrogate model $M_{\theta}$ is employed that substitutes the process input space distribution $\underline{X}$ with the model input space distribution $\underline{\tilde{X}}$ and thus allows for the calculation of $J$.

\begin{algorithm}[H]
	\caption{The RHC-Like Optimization Using Simulated Annealing}
	\textbf{Parameters:} Number of data points $N$, weight coefficients $\underline{q}$, time horizon $L$. \\
	\textbf{Initialization:} Constrained space of the controllable inputs $\mathcal{U}$, constrained model input space $\mathcal{X}$, distance metric dataset $\underline{\Psi}$, surrogate model $\mathcal{M}_\theta$, initial model state  $\underline{\tilde{x}}(0)$, dimension $p$ of the model's input space. \\
	\begin{algorithmic}[H]
		\For{$k = 1, 2, \dots, N$} \Comment{Timesteps}
		\State \begin{align}
			\label{eq:RHClikeOptimization}
			\begin{split}
				\underline{ \hat{U} }_{\mathrm{opt}} & =  \underset{ \underline{\hat{U}} }{ \mathrm{arg\, min} } \ J \bigl(  \underline{\tilde{X}}, \underline{ {\Psi}}, \underline{q} \bigr) \\
				\mathrm{with} \
				& \underline{\tilde{X}} = \mathcal{M}_\theta(\underline{\hat{U}}, \underline{U}, \underline{\tilde{x}}(0)) \\
				\mathrm{s. \, t.\ }  \
				& \underline{\hat{u}}(j) \in \mathcal{U} \hspace{3pt} \quad \forall \, j = \{k, k+1, \ldots, k + L - 1 \} \\ 
				& \underline{{u}}(j) \in \mathcal{U}  \hspace{3pt} \quad \forall \, j = \{1, 2. \ldots, k-1 \} \\ 
				&\underline{\tilde{x}}(j) \in \mathcal{\mathcal{X}}  \hspace{3pt} \quad \forall \, j = \{ 0, 1 \ldots, k, \ldots,  k+L-1 \} \\ 
				& \underline{\psi}(j) \in  \mathbb{R}^p  \quad \forall \, j = \{ 1, 2, \ldots, N_{\Psi} \} \\
				& {q}(j) \in \mathbb{R}  \hspace{7pt} \quad \forall \, j = \{ 1, 2, \ldots, N_{\Psi} \}
			\end{split}
		\end{align}
		\State Apply $\underline{U} \leftarrow \underline{{U}} \cup  \underline{\hat{u}}_{\mathrm{opt}}(k)$. 	\Comment{Append optimal data point}
		\State Optimize $\mathcal{M}_{\theta}$ with data from $\underline{U}$.		\Comment{Only for active learning approach}
		\State Go to $k = k+1$.
		\EndFor
	\end{algorithmic}
\end{algorithm}

\subsubsection*{Novel Optimization Criterion}
The optimization criterion can be mathematically formulated as follows:
\begin{align}
	\label{eq:OptimizationCriterion}
		\begin{split}
		J( \underline{\tilde{X}}, \underline{{\Psi}},  \underline{q} ) &  =  \sum_{j=1}^{N_{\Psi}} q(j) \cdot  d_{\mathrm{NN}}\bigl( {\underline{\psi}}(j), \underline{\tilde{{X}}} \bigr)  \\
		\mathrm{with \, } \ \
		& d_{\mathrm{NN}}\bigl( {\underline{\psi}}(j), \underline{\tilde{{X}}} \bigr)= \underset{1\leq o \leq N_{\underline{\tilde{X}}}}{\min} |\underline{\tilde{x}}(o) - \underline{\psi}(j)| \ .
	\end{split}
\end{align}
In this formulation, $\underline{ {\Psi}}$
represents a distance metric dataset, uniformly distributed within the region of $\underline{\tilde{X}}$, $d_{\mathrm{NN}}$ denotes the nearest neighbor-distance, and $\underline{q}$ contains user-defined weighting coefficients. Hence, $J$ can be interpreted as the weighted sum of the nearest neighbor distances from each point in $\underline{\Psi}$ to $\underline{\tilde{{X}}}$. By adjusting the values within $\underline{q}$, different regions of the model input space can be emphasized with varying intensities.

It is crucial to recognize that minimizing the discrepancies between $\underline{\tilde{X}}$ and $\underline{X}$ is essential for optimal performance. Encouragingly, Heinz et al.$\ $(2017) demonstrated that even a linear time-invariant (LTI) surrogate can produce satisfactory outcomes.  However, for processes characterized by substantial nonlinear behaviors that lead to significant divergences between the LTI and the real input spaces,  a more sophisticated active learning methodology may be employed, wherein $M_{\theta}$ is continually refined using data derived from $\underline{U}$.

\section*{Evaluation}
This section sheds light on the effectiveness of the proposed method in adapting to application-specific requirements by concentrating the information acquisition on process behavior intended to be modeled. Specifically, it is demonstrated how adjusting the weighting coefficients enables targeted emphasis on different regions of the process input space, thereby facilitating a flexible balance between exploration and exploitation.

\begin{figure}[h]
	\centering
	\begin{subfigure}[t]{0.25\textwidth}
		\centering
		\includegraphics[width=\textwidth, trim=3.5cm 11.5cm 3.5cm 11.5cm]{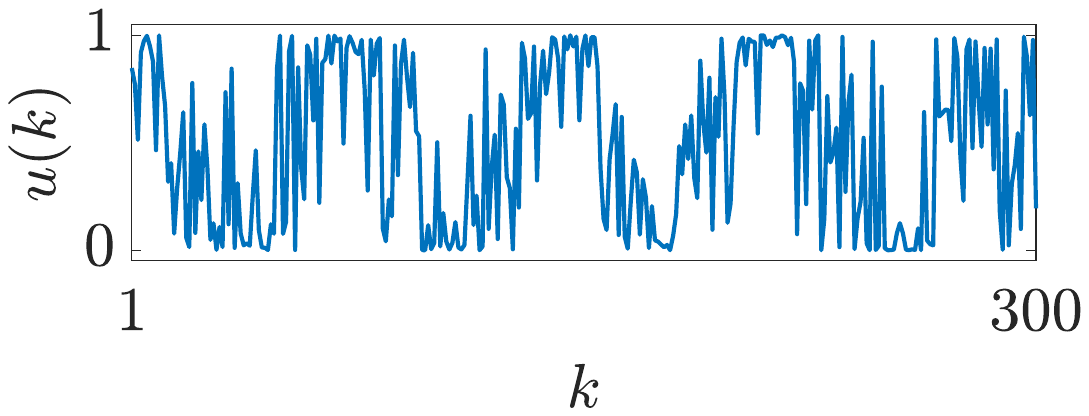}
		\caption{}
		\label{subfig:1}
	\end{subfigure}
\hspace{0.07\textwidth}
	\begin{subfigure}[t]{0.25\textwidth}
		\centering
		\includegraphics[width=\textwidth, trim=3.5cm 11.5cm 3.5cm 11.5cm]{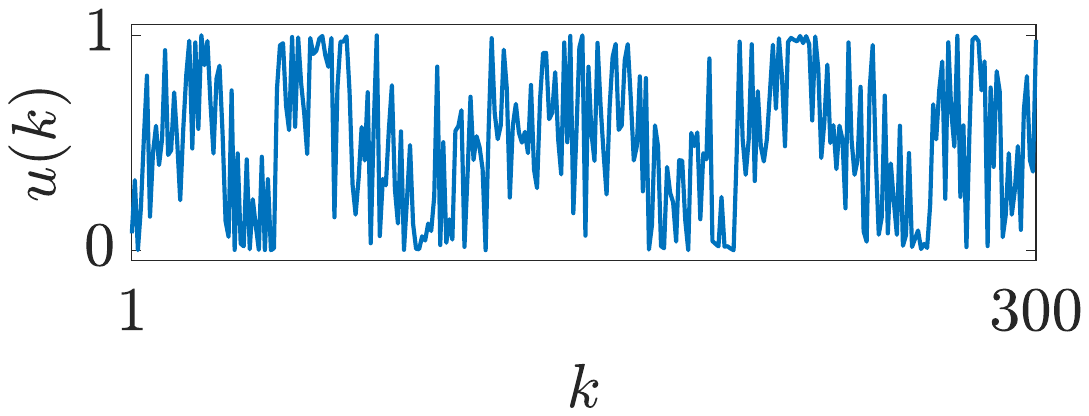}
		\caption{}
		\label{subfig:2}
	\end{subfigure}
\hspace{0.07\textwidth}
	\begin{subfigure}[t]{0.25\textwidth}
		\centering
		\includegraphics[width=\textwidth, trim=3.5cm 11.5cm 3.5cm 11.5cm]{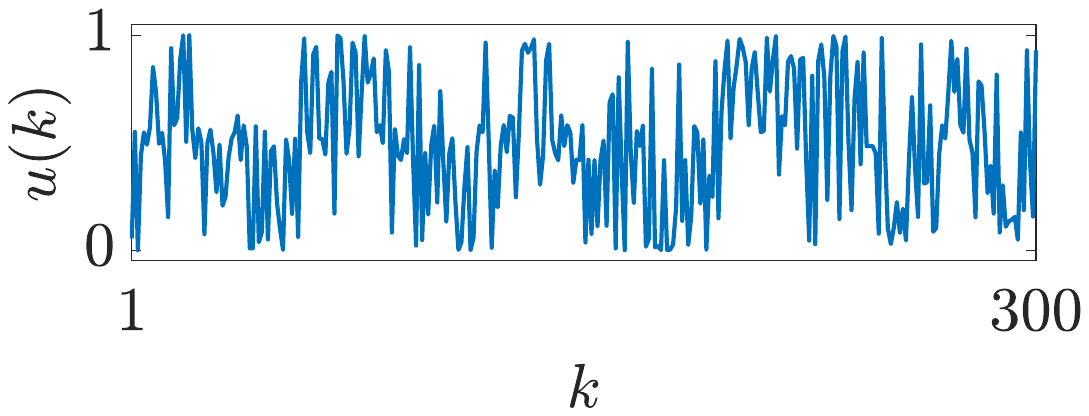}
		\caption{}
		\label{subfig:3}
	\end{subfigure}
	
	\vspace{0.11em} 
	
	\begin{subfigure}[t]{0.25\textwidth}
		\centering
		\includegraphics[width=\textwidth, trim=3.5cm 4.5cm 3cm 4.5cm]{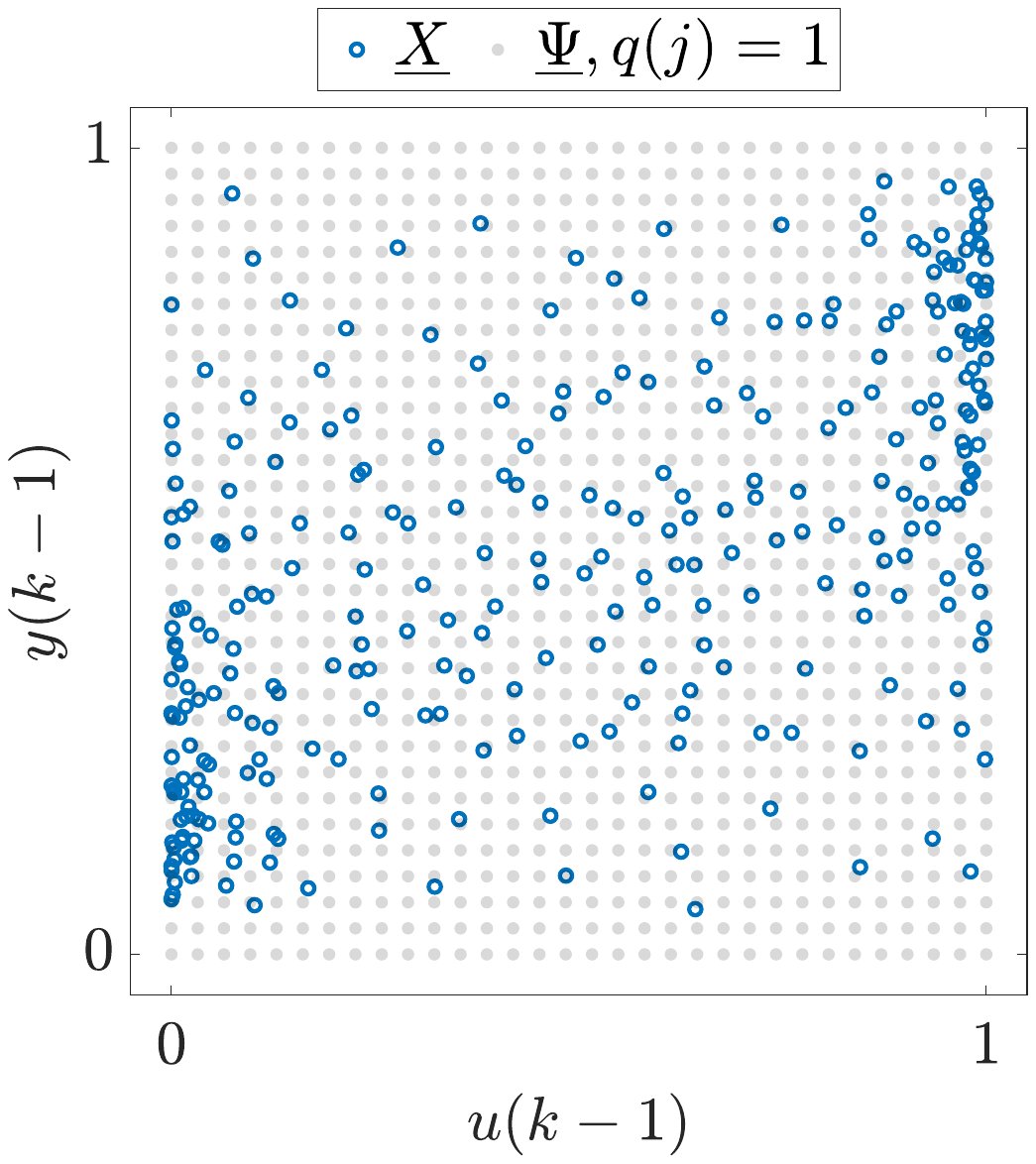}
		\caption{}
		\label{subfig:4}
	\end{subfigure}
\hspace{0.07\textwidth}
	\begin{subfigure}[t]{0.25\textwidth}
		\centering
		\includegraphics[width=\textwidth, trim=3.5cm 4.5cm 3cm 4.5cm]{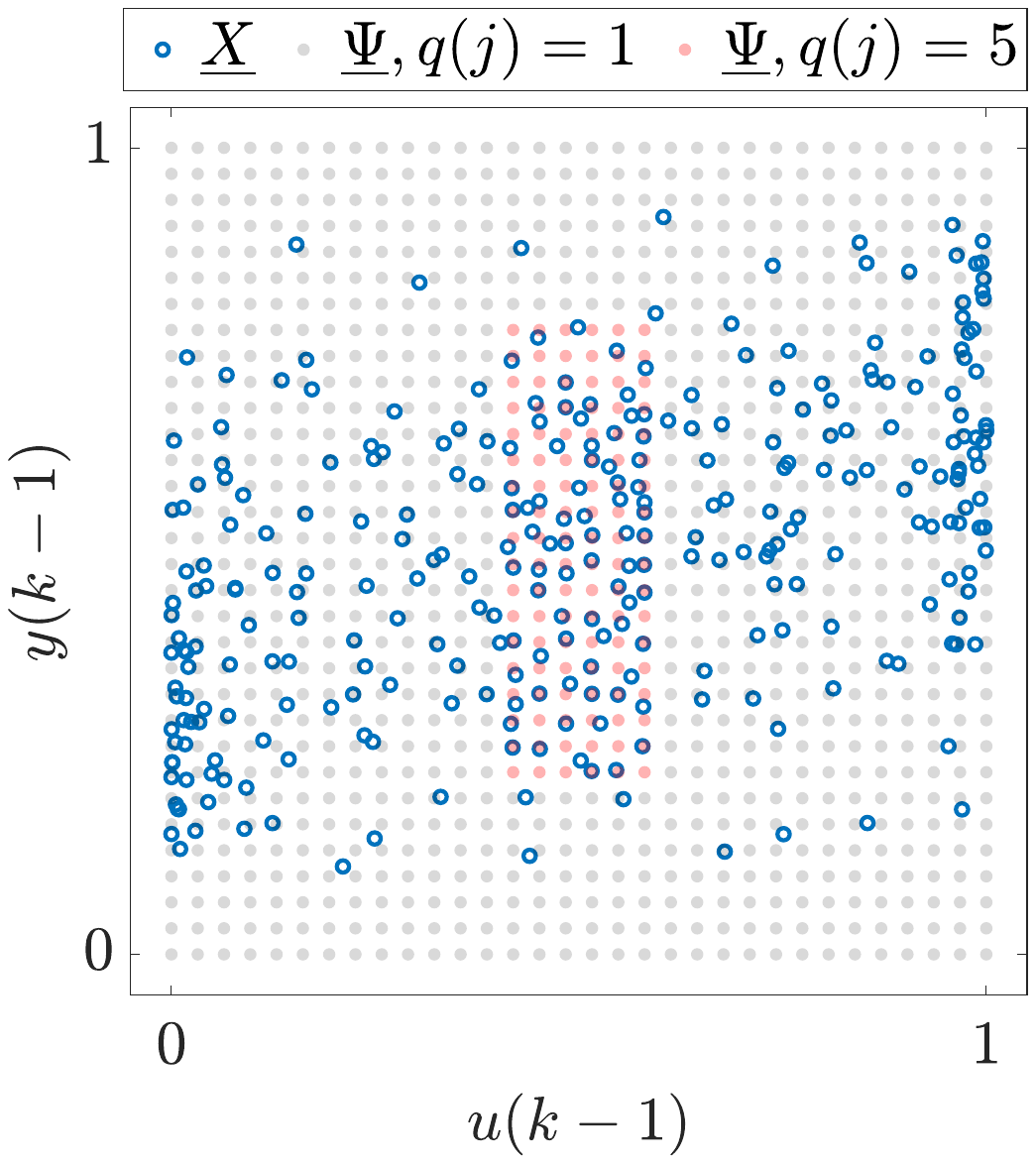}
		\caption{}
		\label{subfig:5}
	\end{subfigure}
\hspace{0.07\textwidth}
	\begin{subfigure}[t]{0.25\textwidth}
		\centering
		\includegraphics[width=\textwidth, trim=3.5cm 4.5cm 3cm 4.5cm]{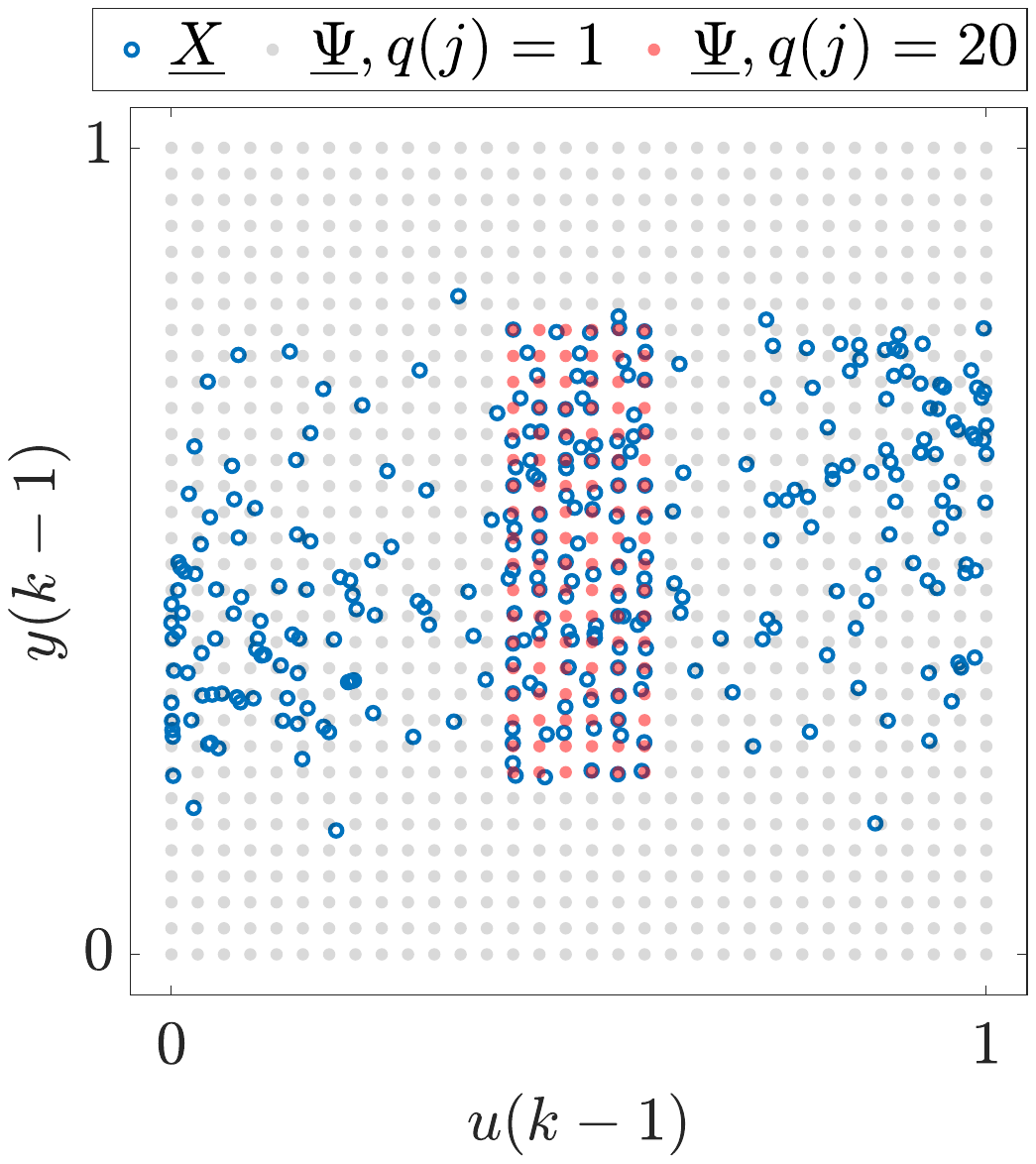}
		\caption{}
		\label{subfig:6}
	\end{subfigure}

	\caption{Excitation signals and below the corresponding process input space distributions employing the example of a nonlinear first-order Hammerstein system. When calculating distances to the red-dotted points of $\underline{ {\Psi}}$, an increased weighting was applied.}
	\label{fig:Signal_and_RSpace}
\end{figure}

Figure \ref{fig:Signal_and_RSpace} illustrates excitation signals and their corresponding distributions in the process input space, generated using distinct weighting schemes. The test process employed is a nonlinear first-order Hammerstein system. Hence, $\underline{x}^{T}(k) = [u(k-1), y(k-1)]$ and $\underline{X} = [\underline{x}(1), \underline{x}(2), \ldots, \underline{x}(N)]$ with $N$ data points. A first-order LTI system is used as surrogate model. In Fig.$\ $\ref{fig:Signal_and_RSpace} (d), an equal weighting of the nearest neighbor distances to each point in $\underline{{\Psi}}$ is applied, resulting in a high-quality space-filling design. This approach is suitable when minimal prior knowledge about the process is available, and the goal is to explore unknown operational regions \cite{heinz2017iterative}. However, if the application requires an intensified information acquisition in regions of higher interest, this can be achieved by increasing the weights of the nearest-neighbor distances calculated to the points of $\underline{{\Psi}}$ in these regions. An example of such targeted exploitation is shown in Fig.$\ $\ref{fig:Signal_and_RSpace} (e) and (f). The progressively increased weighting, illustrated by the red dots, results in a greater concentration of data points in the region of higher interest. This is accompanied by enhanced information acquisition regarding the process behavior in this area.

\section*{Conclusion}
A novel excitation signal design strategy based on a receding horizon control-inspired optimization has been presented. The proposed method has been shown to effectively generate space-filling designs within the input space of a nonlinear dynamic process, thereby enabling sophisticated acquisition of information in previously unexplored operational areas. Additionally, the strategy can intensify the exploitation of specific operational areas during information gathering, offering flexibility in meeting application-specific requirements.


\addtocontents{toc}{\protect\newpage}




\begin{thebibliography}{99}	
	\bibitem{rivera2009constrained}
		Rivera, D.E., Lee, H., Mittelmann, H.D., and Braun, M.W.
		(2009).
		\enquote{Constrained multisine input signals for plant-friendly identification of chemical process systems}.
		In \textit{Journal of Process Control},
		19(4), 
		623-635.
		
		\bibitem{gedlu2023online}
		Gedlu, E.G., Wallscheid, O., Boecker, J., and Nelles, O.
		(2023).
		\enquote{Online system identification and excitation for thermal monitoring of electric machines using machine learning and model predictive control}.
		In \textit{IEEE 14th International Symposium on Diagnostics for Electrical Machines, Power Electronics and Drives (SDEMPED)},
		109-115.
		
		\bibitem{herkersdorf2024idsfid}
		Herkersdorf, M.H., Koesters, T., and Nelles, O.
		(2024).
		\enquote{Optimized Excitation Signal Tailored to Pertinent Dynamic Process Characteristics} [Unpublished Manuscript].
		In \textit{IFAC 4th Modeling, Estimation and Control Conference (MECC)}.
		
		\bibitem{kiss2024space}
		Kiss, M., Toth, R., and Schoukens, M.
		(2024).
		\enquote{Space-Filling Input Design for Nonlinear State-Space Identification}.
		In \textit{arXiv preprint arXiv:2405.18207}.
		
		\bibitem{heinz2017iterative}
		Heinz, T.O. and Nelles, O.
		(2017).
		\enquote{Iterative excitation signal design for nonlinear dynamic black-box models}.
		In \textit{Procedia computer science},
		112, 
		1054-1061.
		
		\bibitem{van1992stochastic}
		Van Kampen, N.G.
		(1992).
		\enquote{Stochastic Processes in Physics and Chemistry}.
		North-Holland Publishing.
		
		\bibitem{kosters2022optimization}
		Koesters, T., Heinz, T.O., and Nelles, O.
		(2022).
		\enquote{Optimization based excitation signal design tailored to application specific requirements}.
		In \textit{IFAC-PapersOnLine},
		55(37)
		451-456.
		
\end{thebibliography}
\end{document}